\documentclass[%
 sor,
 jor,
 amsmath,amssymb,
 reprint,%
]{revtex4-2}

\usepackage{graphicx}
\usepackage{dcolumn}
\usepackage{bm}

\usepackage[margin=2cm]{geometry}
\usepackage[most]{tcolorbox}

\DeclareMathOperator*{\cov}{Cov} 
\DeclareMathOperator*{\var}{Var}	

\newcommand{\PCCodf}{\rho_{d,f}} 
\newcommand{\SDf}{\sigma_{f}} 
\newcommand{\SDod}{\sigma_{d}} 

\usepackage{color}

\newtcolorbox{mytextbox}[1][]{%
  sharp corners,
  enhanced,
  colback=white,
  height=3cm,
  attach title to upper,
  #1
}

\begin{document}

\begin{mytextbox}
This is a draft of a chapter. The final version will be available in the Handbook of Computational Social Science edited by Taha Yasseri, forthcoming 2025, Edward Elgar Publishing Ltd. The material cannot be used for any other purpose without further permission of the publisher and is for private use only. 

Please cite as: Kristina Lerman (2025). Strong Friendship Paradox in Social Networks. In: T. Yasseri (Ed.), Handbook of Computational Social Science. Edward Elgar Publishing Ltd.
\end{mytextbox}

\title{Strong Friendship Paradox in Social Networks}

\author{Kristina Lerman}
 \altaffiliation[]{USC Information Sciences Institute}

\date{\today}

\begin{abstract}
The friendship paradox in social networks states that your friends have more friends than you do, on average. Recently, a stronger variant of the paradox was shown to hold for most people within a network: “most of your friends have more friends than you do.” Unlike the original paradox, which arises trivially because a few very popular people appear in the social circles of many others and skew their average friend popularity, the strong friendship paradox depends on features of higher-order network structures. Similar to the original paradox, the strong friendship paradox generalizes beyond popularity. When individuals have traits, many will observe that most of their friends have more of that trait than they do. This can lead to the “majority illusion,” in which a rare trait will appear highly prevalent within a network. Understanding how the strong friendship paradox biases local observations within networks can inform better measurements of network structure and our understanding of collective phenomena in networks.

\end{abstract}

\maketitle

\section{Introduction}
Social networks represent the interpersonal ties that bind people by friendship, membership in a community, work collaboration, and other social interactions. These relationships play an important role in shaping individual beliefs and behaviours, but also collective dynamics of social networks, including the emergence of social norms and consensus in a population, segregation along demographic and socio-economic lines, and affective polarization and the emergence of echo chambers in online social networks~\cite{Schelling73,Granovetter78,hogg2009stochastic,Centola10}. 


The field of network science studies the interplay between the structure of social networks and individual behaviour, and how that affects collective outcomes in networks. Consider the social network shown in Figure~\ref{fig:SFP}. This classic network shows friendships (edges) between members (nodes) of a university karate club~\cite{zachary1977information}. Though this network is orders of magnitude smaller than what network scientists now study, its structure is typical of other social networks. For example, some nodes have many more edges than others. We refer to this quantity as a node's degree. Also, the network's community structure is clearly visible, with two dense clusters centred      around two high degree nodes.

\begin{figure}
    \centering
    \includegraphics[width=0.4\linewidth]{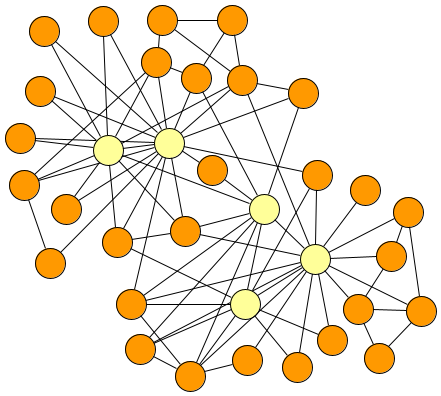} 
    \caption{Friendship paradox in a social network representing friendship relationships between members of a university karate club~\protect\cite{zachary1977information}. Darker nodes represent members who observe that they are less popular (have fewer friends) than their friends are on average.}
    \label{fig:FP}
\end{figure}

People rarely get a global view of the social network; instead, they must infer its state from their local observations of their friends. However, friends are often systematically different from others in the social network. Since the behaviours of individuals are shaped by their perceptions of what others are doing, this difference can systematically distort collective behaviour in social networks. One source of bias in individual perceptions is the \textit{friendship paradox}, which states that ``your friends have more friends than you do, on average''~\cite{Feld91}. Stated more precisely, given a social network, the degree of a randomly chosen node tends to be smaller than the average degree of its friends. Getting back to the friendship network in Fig.~\ref{fig:FP}, the darker nodes all experience the friendship paradox.

The friendship paradox, while innocuous on the surface, leads to a surprising array of phenomena in social networks. One of these is the \textit{generalized friendship paradox}, which states that friends have more of some trait: friends are happier, on average~\cite{Bollen2017}, and collaborators produce more impactful research~\cite{Benevenuto2016}. The friendship paradox explains why people overestimate the prevalence of risky behaviours among their peers~\cite{Baer91,berkowitz2005overview}, the size of minority groups in a community~\cite{Lee2019}, and the popularity of topics on social media~\cite{alipourfard2020friendship}. Researchers have used the friendship paradox for early detection of flu outbreaks~\cite{christakis2010}, to predict what topics will become popular on social media~\cite{GarciaHerranz14}, and to efficiently poll public opinion~\cite{nettasinghe2018your}.


The friendship paradox is a straightforward consequence of sampling from a heterogeneous distribution. Such distributions describe populations with a high degree of variability or diversity. For example, in a network with a hetereogeneous degree distribution, outliers with many friends are not that rare. As long as there is a variance of degree, a popular friend will appear in many social circles, skewing the mean friend count for many. To reduce the impact of outliers, Kooti et al.,~\cite{Kooti14icwsm} introduced the \textit{strong friendship paradox}, which compares an individual's degree to the median of their friends’ degrees, rather than the mean. Strong friendship paradox states that ``most of your friends are more popular than you are.'' Surprisingly, this stricter comparison holds for the vast majority of nodes in real-world networks. Unlike the canonical friendship paradox that can be explained by the shape of the degree distribution, the stronger version requires deeper knowledge of network structure, specifically, the property of networks called transsortativity, which measures the correlation of degrees of node's neighbours~\cite{ngo2020transsortative}. Knowing transsortativity is necessary to accurately estimate how many nodes experience the strong friendship paradox~\cite{Wu2017neighbor}. 

The strong friendship paradox acts like a funhouse mirror to distort reality for individuals by exaggerating the prevalence of a trait. It can lead to the ``majority illusion''~\cite{Lerman2016majority}, where a rare trait can seem extremely popular, appearing within many social circles. In complex contagion (see~\cite{Centola10,Centola15}), where a node adopts a behaviour when some fraction of friends (e.g., a majority) adopt it, the majority illusion could trigger a global spread from a vanishingly small number of seeds. 

In this chapter, we survey existing research to situate strong friendship paradox within the context of the friendship paradox and its variants. This helps to surface knowledge gaps in our understanding of network biases and identify new research directions.

\section{Taxonomy of Paradoxes in Networks}
\subsection{Friendship Paradox}
\label{sec:fp}
First, we review the traditional friendship paradox, its generalizations to attributes other than degree and link it to perception bias.

\subsubsection{Friendship Paradox for Degree}
The canonical statement of the friendship paradox (FP)---``your friends have more friends than you do, on average''---has been interpreted by researchers in different ways~\cite{Feld91,cantwell2021friendship}. 
Following \cite{nettasinghe2019friendship,alipourfard2020friendship}, we consider a network $G = (V,E)$, with $\{V\}$ nodes and $\{E\}$ undirected links, and define a random variable $X$ as a \textit{random node} sampled uniformly from  $\{V\}$, and random variable $Y$ as a \textit{random friend} sampled from $\{V\}$ proportional to its degree. Friendship paradox can then be stated as follows: the average degree of a random friend ($\mathbb{E}\{d(Y)\}$) is larger than average degree of a random node ($\mathbb{E}\{d(X)\}$). This inequality holds under the condition that variance of degree distribution is greater than zero~\cite{nettasinghe2019friendship}. 
\begin{equation}
\mathbb{E}\{d(Y)\} - \mathbb{E}\{d(X)\} = \frac{\var\{d(X)\}}{\mathbb{E}\{d(X)\}} \geq 0.
\end{equation}

Friendship paradox extends to structural properties in network other than degree: friends are not only more popular, but also more important, in a sense that they have higher network centrality \cite{higham2019centrality}.


The definition has been extended to directed networks, such as the follower graphs on social media platforms, where a directed edge from $u$ to $v$ signifies that $u$ follows $v$, e.g., receives updates from $v$, but not necessarily vice versa. This changes the notion of degree. We refer to nodes that $u$ follows as its \textit{friends} (out-degree), and nodes that follow it as \textit{followers} (in-degree). Four distinct versions of FP exist in directed networks, and vast majority of users of social media platforms like Digg and Twitter experience these paradoxes~\cite{Hodas13icwsm}. Two of the paradoxes---namely 1) your friends have more followers than you do, on average, and 2) your followers have more friends than you do, on average---are similar to the canonical FP in that they require variance in the distribution of the number of friends and followers, respectively. However, the remaining two paradoxes---3) your friends have more friends than you do, on average, and 4) your followers have more follower than you do, on average---require an additional network property to be satisfied, specifically they require a positive correlation between the number of friends and followers of nodes in the network~\cite{alipourfard2020friendship}. 

Friendship paradox also has useful applications in disease monitoring, polling, and trend prediction. It is the foundation for the formation of social norms~\cite{jackson2019friendship}, and explains distortions in the perception of social norms, including why college students overestimate how much their peers engage in risky behaviors, such as binge drinking and substance abuse~\cite{Baer91,berkowitz2005overview}.
Researchers have used the friendship paradox for early detection of flu outbreaks on a college campus~\cite{christakis2010} and to devise efficient polling strategies to estimate public opinion~\cite{nettasinghe2018your} and to predict what topics will be popular on social media~\cite{GarciaHerranz14}. The basic idea behind FP-based surveillance is that rather than monitor a few randomly-chosen individuals, for example, to see if they have caught the flu, you instead ask them who their friends are and monitor a similar number of randomly-chosen friends. This policy creates more precise surveillance for the same monitoring cost: friends are more central in the friendship network; therefore, more likely to become infected first with a pathogen or a contagious idea.

\subsubsection{Generalized Friendship Paradox}
When nodes have distinguishing attributes, the paradox can distort their perceptions of how prevalent those attributes are among friends. For simplicity, we assume that each node has a binary attribute  ($f: V \rightarrow \{0,1\}$), which can represent gender (female or male), partisanship (liberal vs conservative), affective state (happy vs unhappy), or behavior (bought a product or not).
The prevalence of the attribute in a network is given by $\mathbb{E}\{f(X)\}$, the expectation a random node $X$ has the attribute. For example, when 15\% of people have a new iPhone ($f(v)=1$), $\mathbb{E}\{f(X)\}=0.15$.  The \textit{perceived} prevalence of the attribute, however, can be very difference under some conditions. The perceived prevalence is given by the expected value of the attribute among friends  $\mathbb{E}\{f(Y)\}$.  This leads to the \textit{generalized friendship paradox}, which states that \textit{your friends have more of the attribute that you do, on average}. 
Your friends are more active on social media platforms~\cite{Hodas13icwsm}, happier~\cite{Bollen2017}, and see more novel and viral content~\cite{Kooti14icwsm}. Your collaborators have higher h-index~\cite{Benevenuto2016} and are more productive~\cite{Eom14_2} than you are.

The generalized friendship paradox exists whenever high-degree nodes are also more likely to have the trait~\cite{momeni2015measuring,Eom14,fotouhi2014generalized}:

		\begin{align}
		  \label{eq:global_perception_bias_Y}
			\mathbb{E}\{f(Y)\} - \mathbb{E}\{f(X)\} &= \frac{\cov(f(X),d(X))}{\mathbb{E}\{d(X)\}} 
			=\frac{\PCCodf\SDod \SDf}{\mathbb{E}\{d(X)\}},	
		\end{align}

\noindent where $\PCCodf$ is the Pearson correlation coefficient between degree and attribute value of a random node, $\SDod$ is the standard deviation of the degree distribution, and $\SDf$ is the standard deviation of the binary attributes~\cite{alipourfard2020friendship}.

\subsubsection{Perception Bias}
Rather than compare themselves to their friends, nodes can use their observations of friends to estimate the prevalence of some trait, for example, a controversial opinion or religious membership. FP can systematically skew these observations, creating a \textit{perception bias} in networks. Lee et al.~\cite{Lee2019} showed that in networks with a majority and a minority group, nodes can overestimate the size of the minority group, and this mis-estimation 
depends on the size of the group and also the level of homophily in the network, where homophily quantifies node's propensity to connect to similar nodes, e.g., minority nodes linking to other minority nodes. 

In a study of popular topics on social media, Alipourfard et al.~\cite{alipourfard2020friendship} identified global and local perception bias, which leads nodes to systematically overestimate (or underestimate) the popularity of some topic. Similar to the generalized FP, \textit{global perception bias} depends on correlation of node's degree and its attribute. In other words, when popular users post about some topic, it will appear as if there were more people  discussing that topic than there really are. Alipourfard et al. also defined local perception bias, which measures the fraction of friends with the trait (e.g., talking about the topic). They identify topics that appear more often among the posts users see within their social feeds than they do globally among all posts. They showed that local perception bias can make some viral topics appear up to four times more popular than they really are. For local perception bias to exist, two conditions need to be satisfied: first, as with global perception bias, high-degree nodes must be more likely to have the trait, but also they must be followed by low-attention users who do not follow many other people. Interestingly, though FP biases individual node's estimates of the prevalence of some trait or discussion topic, it can be leveraged within a polling algorithm to obtain a statistically efficient estimate of a topic’s global prevalence from biased individual perceptions~\cite{alipourfard2020friendship}.

\subsection{Strong Friendship Paradox}
\label{sec:SFP}
We introduce strong friendship paradox and its generalization to other attributes than degree, and also show its relationship to the majority illusion in social networks.
\subsubsection{Strong Friendship Paradox for Degree}
\begin{figure}
    \centering  
    \includegraphics[width=0.4\linewidth]{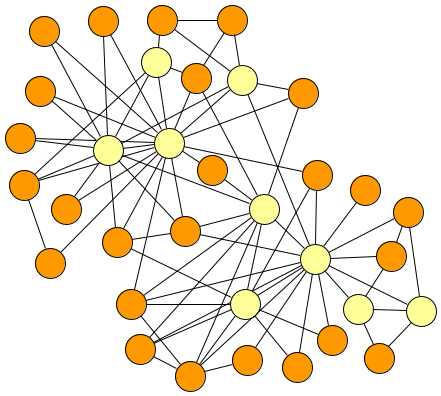} 
    \includegraphics[width=0.4\linewidth]{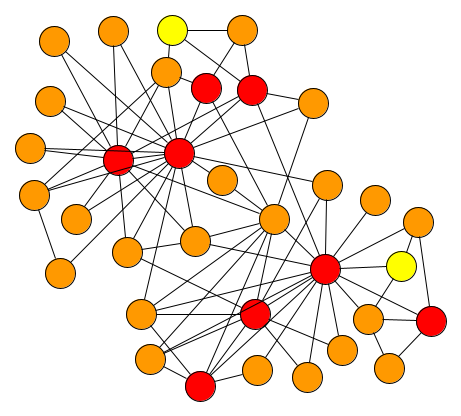}
    \caption{Illustrations of network paradoxes in the karate club social network. 
    (left) \textit{Strong friendship paradox}: Darker nodes represent members that observe that they have fewer friends than most of their friends do.
    (right) \textit{Majority illusion}: Although only eight of the 34 nodes are red, all but two of the remaining nodes (yellow) observe that at least half of their friends are red. }
    \label{fig:SFP}
\end{figure}

Instead of comparing a node's popularity to friends' mean popularity, Kooti et al.~\cite{Kooti14icwsm} proposed comparing it to the median. This change leads to what we call \textit{strong friendship paradox}, which states that ``most of your friends have more friends than you do.'' 
Though it sounds even more paradoxical than the canonical FP, strong friendship paradox affects large number of node---between 70 and 90\% of all nodes in a variety of empirical social networks~\cite{Wu2017neighbor} (see Fig.~\ref{fig:hep_SFP}(b)). Figure~\ref{fig:SFP} illustrates SFP on a benchmark social network of Zachary's karate club~\cite{zachary1977information}. The orange nodes experience SFP: most of their friends have at least as many friends as they do.

\subsubsection{Generalized Strong Friendship Paradox}
Strong friendship paradox generalizes to any trait or attribute that is correlated with degree. Kooti et al. found that most of user's friends on social media are more active, i.e., they post more content than the user does. Most of the friends also receive more diverse and viral information than the user sees in her own feed.

\subsubsection{Majority Illusion}

The strong friendship paradox can dramatically distort a node's  local neighborhood, leading to the ``majority illusion''~\cite{Lerman2016majority} in which a globally rare trait may be overrepresented among the friends of many nodes. To see how, consider again the karate club social network shown in Fig.~\ref{fig:SFP}(right). Suppose that red nodes are members who hold some unpopular belief or a trait. Although a small fraction of the nodes are red (eight of 34), all but two of the remaining nodes (yellow nodes) observe that at least half of their friends are red. This means that this rare trait will appear to be popular locally. In complex contagions, where the trait spreads in a social network after nodes observe enough of their friends with the trait,  a small minority can cause the trait or belief to spread globally. Thus, the majority illusion can dramatically amplify exposure to some rare information in networks, affecting collective response and network processes such as dynamics of contagion.

Mathematical analysis of the majority illusion~\cite{Lerman2016majority}  shows that magnitude of the effect depends on the degree-attribute correlation and it is larger when the more popular nodes hold the rare trait. Moreover, the effect is larger in disassortative networks where popular nodes tend to connect to less popular nodes.

\section{Mathematical Analysis}

Canonical FP is rooted in a heterogeneous degree distribution: it exists in any network with a variance in its degree distribution. In such networks, there exists individuals who are unexpectedly popular, and they skew the average friend popularity of many others simply because they appear within the social circles of many others individuals. Stated this way, FP (and its generalizations) does not require a network, but simply arises as a generic property of sampling from skewed distributions, as described below.

More sophisticated mathematical analysis of the FP shows that to estimate its strength, i.e., probability that a node in a given network experiences FP, one needs to take into account degree correlations of linked nodes~\cite{cantwell2021friendship,pal2018quantifying}, a quantity known as degree assortativity in network science~\cite{newman2002assortative}. 

Unlike the canonical friendship paradox, the strong friendship paradox cannot be explained by statistical sampling and requires knowledge about network structure.  

\subsection{Statistical Explanation of Friendship Paradox}

Friendship paradox is an artifact of sampling from skewed distributions. To understand, consider a related phenomenon called the \textit{class size paradox}~\cite{Feld91}. This paradox explains why students generally experience classes as being larger than they typically are, and why motorists experience freeways as being more congested than they really are. The paradox arises because large classes (resp. jammed freeways) by definition have many students (resp. motorists), whose observations skew the estimates of the mean class size (resp. congestion).  

Similar reasoning explains the generalized FP. Consider a random network where nodes have traits $X$ that are i.i.d. random variables from a heterogeneous distribution. The generalized FP compares node $i$th value of the trait $x_i$ to the mean value of node's $k$ neighbors $\frac{1}{k} \sum_{j=1}^k x_j$. We know that the mean of a set of $k$ numbers drawn from a heterogeneous distribution grows with $k$; therefore, mean of $k$ numbers is almost always larger than a single value $x$. A more rigorous statistical argument can be found in Kooti et al.~\cite{Kooti14icwsm}.

\subsection{Network Structure in Strong Friendship Paradox}

In contrast to the canonical FP, strong friendship paradox does not arise purely from considerations of statistical sampling. A shuffling test that scrambles the values of the trait $x$ among network nodes thereby destroying correlation between the trait and degree, eliminates the strong FP but not the canonical FP~\cite{Kooti14icwsm}. This is because shuffling preserves the skewed distribution of the trait, and generalized FP existence only depends on this distribution.

\begin{figure}
    \centering

\begin{tabular}{cc}
    \begin{minipage}[c]{0.5\textwidth}
    \includegraphics[width=\linewidth]{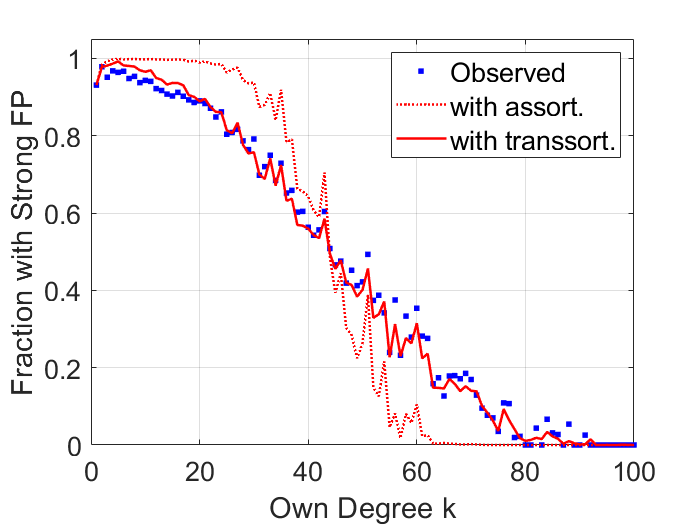}
    \end{minipage}
        &
\begin{minipage}[c]{0.5\textwidth}
\begin{tabular}{|l|l|c|c|}\hline
\textit{Network} & \textit{Type} & \textit{Observed} & \textit{Predicted} \\\hline\hline
LiveJournal & \small{Social} & 83.71\% & 84.43\%\\\hline
Youtube & \small{Social} & 89.94\% & 88.51\%\\\hline
Skitter & \small{Internet} & 88.62\% & 90.42\%\\\hline
Google & \small{Web} & 77.31\% & 78.43\%\\\hline
ArXiv HEP & \small{Citation} & 78.71\% & 79.76\%\\\hline
English words & \small{Semantic} & 75.23\% & 71.21\% \\\hline
\end{tabular}
\end{minipage}
\\
(a) & (b)
    
\end{tabular}

    \caption{Strong FP in real-world networks (data from \protect\cite{Wu2017neighbor}). (a) The plot shows the fraction of nodes in a citation network of physics papers with $k$ citations that experience the strong friendship paradox. To experience the paradox, a node with degree $k$ must observe that most of its neighbors have degree at least $k$. The lines show predictions of the model that includes degree distribution and degree assortativity (dashed), and also transsortativity (solid) capturing degree correlations between node's neighbors. Both models and observational measurements ignore direction of edges and consider as neighbors both cited and citing papers.
    (b) Observed fraction of nodes in real-world networks that experience the strong friendship paradox, compared to predicted prevalence of the paradox with the transsortativity mode.}
    \label{fig:hep_SFP}
\end{figure}

To explain strong friendship paradox, more information about the network is needed.
Wu el al.~\cite{Wu2017neighbor} described a mathematical model to estimate the magnitude of SFP in a given network, i.e., the probability a node of degree $k$ will experience the paradox. They showed that SFP is more prevalent in disassortative networks, where high degree nodes tend to connect to low degree nodes. However, accurate estimation of the paradox also requires knowing degree correlations of node's neighbors. This property, called transsortativity~\cite{ngo2020transsortative}, can significantly alter network structure and network phenomena, including the ``majority illusion'' and the critical threshold for cascades in the Watts threshold model of contagion~\cite{wu2018degree}.

Figure~\ref{fig:hep_SFP} quantifies the strength of the paradox. Fig.~\ref{fig:hep_SFP}(a) shows the probability a node of degree $k$ will experience the strong friendship paradox, along with analytic estimate of the paradox strength. The model that takes transsortativity into account (solid line) makes the more accurate predictions. The table in Fig.~\ref{fig:hep_SFP}(b) shows the prevalence of the SFP in several real-world networks and estimates its strength.

\section{Conclusion}
The connection between biased local views and network structure revealed by the strong friendship paradox is important for several reasons. First, it may not always be practical to observe large networks in their entirety: instead, we can sample nodes and estimate network properties by exploring their local neighbourhoods. We must, therefore, be aware of how the paradox may systematically bias local measurements of network structure, including sampled degree distribution. 
     
Furthermore, the strong friendship paradox creates a perception bias that affects how information is interpreted within networks. In real-world social networks, this bias can amplify certain beliefs and behaviors. High-degree influencers, for instance, exert disproportionate influence on the network because their behaviors and opinions are more frequently observed and emulated. When multiple high-degree influencers share a common trait, it can appear far more prevalent than it actually is, further distorting perceptions of social norms. This distortion may lead individuals to overestimate the prevalence of certain behaviours within their networks, such as illicit substance use or extreme fitness trends. Perceived norms, shaped by these biases, can create social pressure, accelerating the adoption of behaviours—including harmful ones like substance abuse or eating disorders.

To mitigate the effects of these distorted perceptions, it is essential to account for the strong friendship paradox and its interaction with higher-order network structures. Doing so can improve our understanding of network dynamics and enable the design of interventions to counteract the amplification of harmful behaviours and misperceptions.

\begin{acknowledgments}
This work was supported in part by the Air Force Office of Scientific Research under contract FA9550-20-1-0224. The author is grateful to Petter Holme for providing valuable feedback.
\end{acknowledgments}



\bibliography{references}

\end{document}